\documentclass[structabstract]{aa}  
\usepackage{lscape}
\usepackage{amsmath}
\usepackage{graphicx}
\usepackage{natbib}
\usepackage{txfonts}
\usepackage{longtable}
\usepackage{amsfonts}

\newcommand{\parameters}{
\begin{table}[t!]
\caption[]{Characterizing parameters of HD\,95881. The second Col.
lists parameters as given by  \cite{2004A&A...426..151A}, where SFR
stands for the star formation region. New values for some of the 
parameters are given in the third Col.}
\label{tab:parameters} 
\centering   
\begin{tabular}{lcc} 
\hline\hline\\[-9pt]   
Parameter             & Value                                & New         \\  
\hline\\[-9pt]                          
Right Ascension       & 11$^{\rm h}$ 01$^{\rm m}$ 57\,$\fs$62 &             \\
Declination           & -72$^\circ$ 30$\arcmin$ 48$\farcs$4  &             \\
Spectral Type         & A2III/IVe                            &             \\
$T_{\rm eff}$\,[K]    & 8990                                 &             \\
Distance\,[pc]        & 118$\pm12$                           & 170$\pm$30  \\
Luminosity\,[L$_{\odot}$] & 6.9$\pm$1.0                      & 15.4$\pm$6  \\
$A_{\rm V}$ [mag]     & 0.25                                 &             \\
Radius\,[R$_\odot$]   & 1.1$\pm$0.1                          & 1.6$\pm$0.3 \\
Mass\,[M$_{\odot}$]   & -                                    & 2.0$\pm$0.3 \\
SFR                   & Sco OB2-4?                           & ?           \\
Group                 & IIa                                  &             \\
\hline
\end{tabular}
\end{table}
}

\newcommand{\modelfit}{
\begin{table}[t!]
\caption{The parameters for our final model for the circumstellar 
disk of HD\,95881. \ch{The carbon abundance was found to be 25\% of
the visible dust.} The \ch{visible} dust grains and the PAH molecules
have a different spatial distribution.  }
\begin{center}
\begin{tabular}{lcc}
\hline\hline\\[-9pt]
Parameter   & Dust & PAHs\\
\hline\\[-9pt]  
Inner disk radius ($R_\mathrm{in}$)		& 0.55\,AU                & 0.55\,AU \\
Outer disk radius ($R_\mathrm{out}$)		& 200\,AU                 & 200\,AU  \\
Inclination angle ($i$)                         & 55$^\circ$              & 55$^\circ$ \\
Position angle (major axis E of N)		& 103$^\circ$             & 103$^\circ$ \\
Vertical density scaling parameter ($\Psi$)	& 2.75                    & 2.75 \\
Power law for the surface density ($p$)		& 1                       & 1   \\
Turnover point ($R_0$)				& 2.5\,AU                 & $\infty$ \\
Mass ($M$)			                & 10$^{-8}$\,M$_{\odot}$  & 5$\cdot$10$^{-9}$\,M$_{\odot}$ \\
\hline
\end{tabular}
\end{center}
\label{tab:modelfit}
\end{table}
}

\newcommand{\amber}{
\begin{figure}[t!]
\centering
\includegraphics[width=\columnwidth]{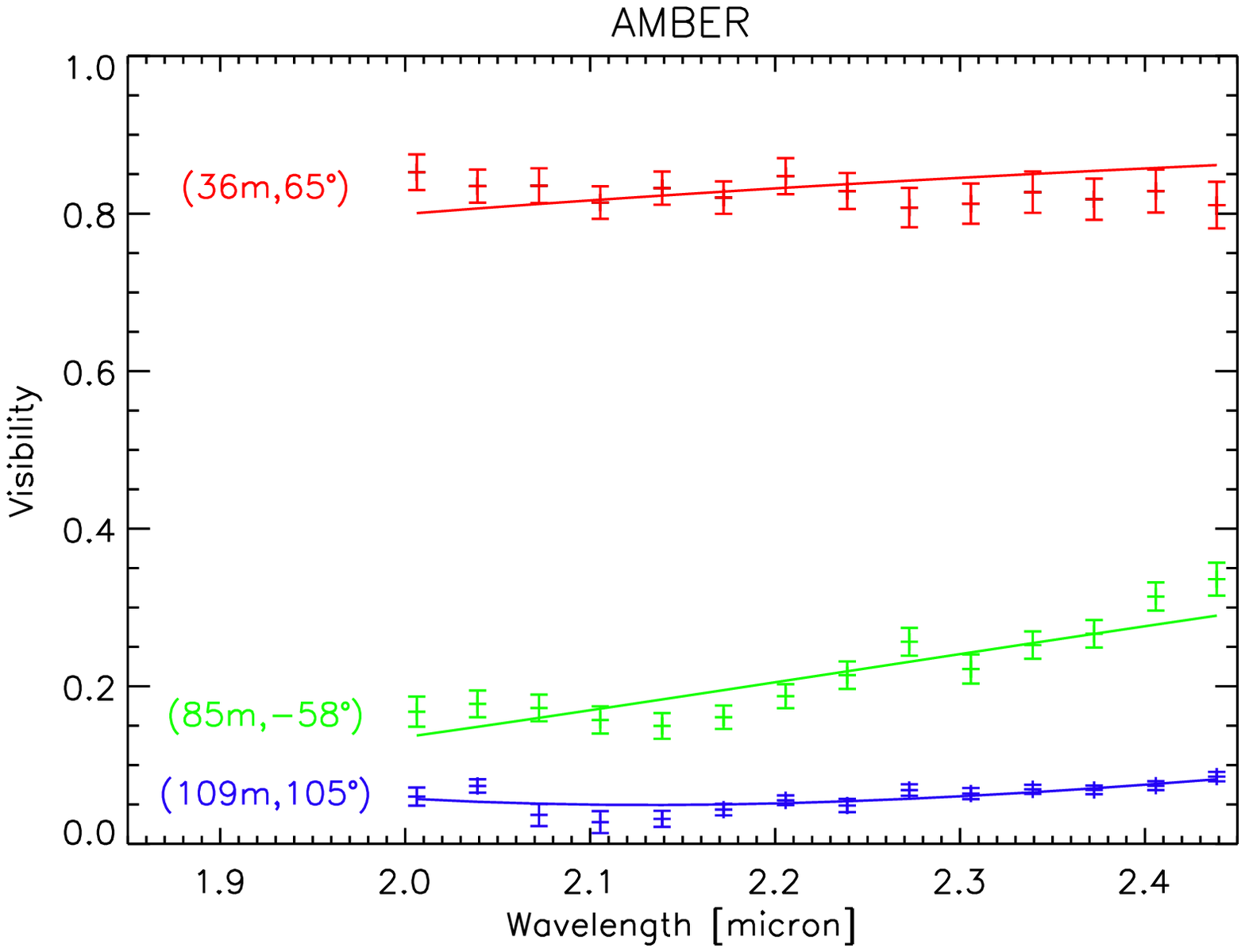}
\caption{The AMBER K-band visibilities of HD\,95881. The projected
length and position angle of the baselines is indicated on the
left. Overplotted is the best-fit ring+point-source model (solid
lines; see text).  }
\label{fig:amber}
\end{figure}
}

\newcommand{\profile}{
\begin{figure}[t!]
\centering
\includegraphics[width=\columnwidth]{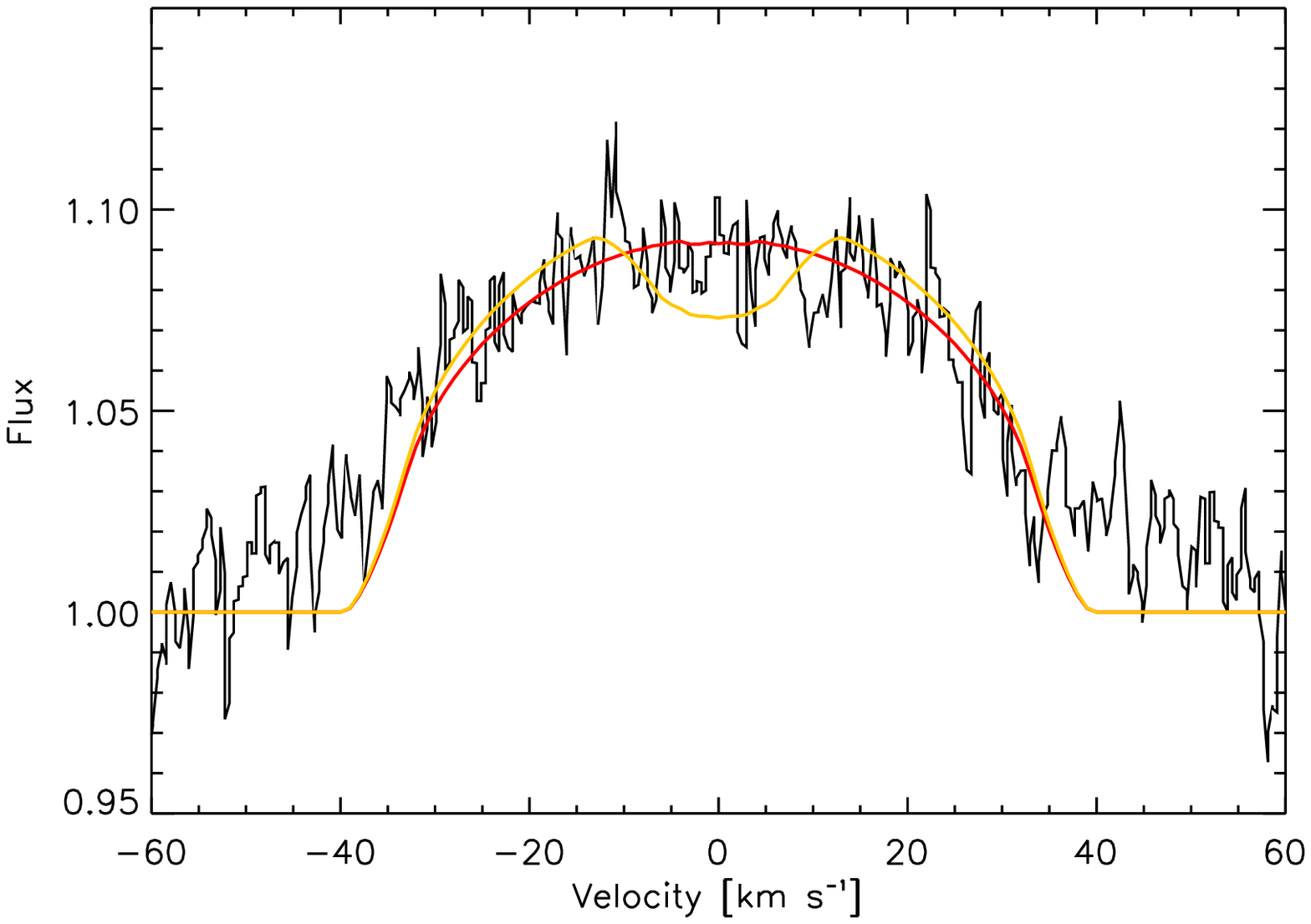}
\caption{The continuum-normalized [O\,I]\,6300\,\r{A} emission line of
HD\,95881 (\citealt {2005A&A...436..209A}). The velocity axis has been
centered on the centroid position of the feature, i.e. the radial
velocity of the central star. Line profiles corresponding to a surface
brightness proportional to $r^{-3}$ are overplotted. Red represents a
model with an outer radius beyond 31\,AU; yellow \ch{corresponds to}
the model with an outer radius equal to 11\,AU. See text for details.}
\label{fig:profile}
\end{figure}
}

\newcommand{\spitzer}{
\begin{figure}[t!]
\centering
\includegraphics[width=\columnwidth]{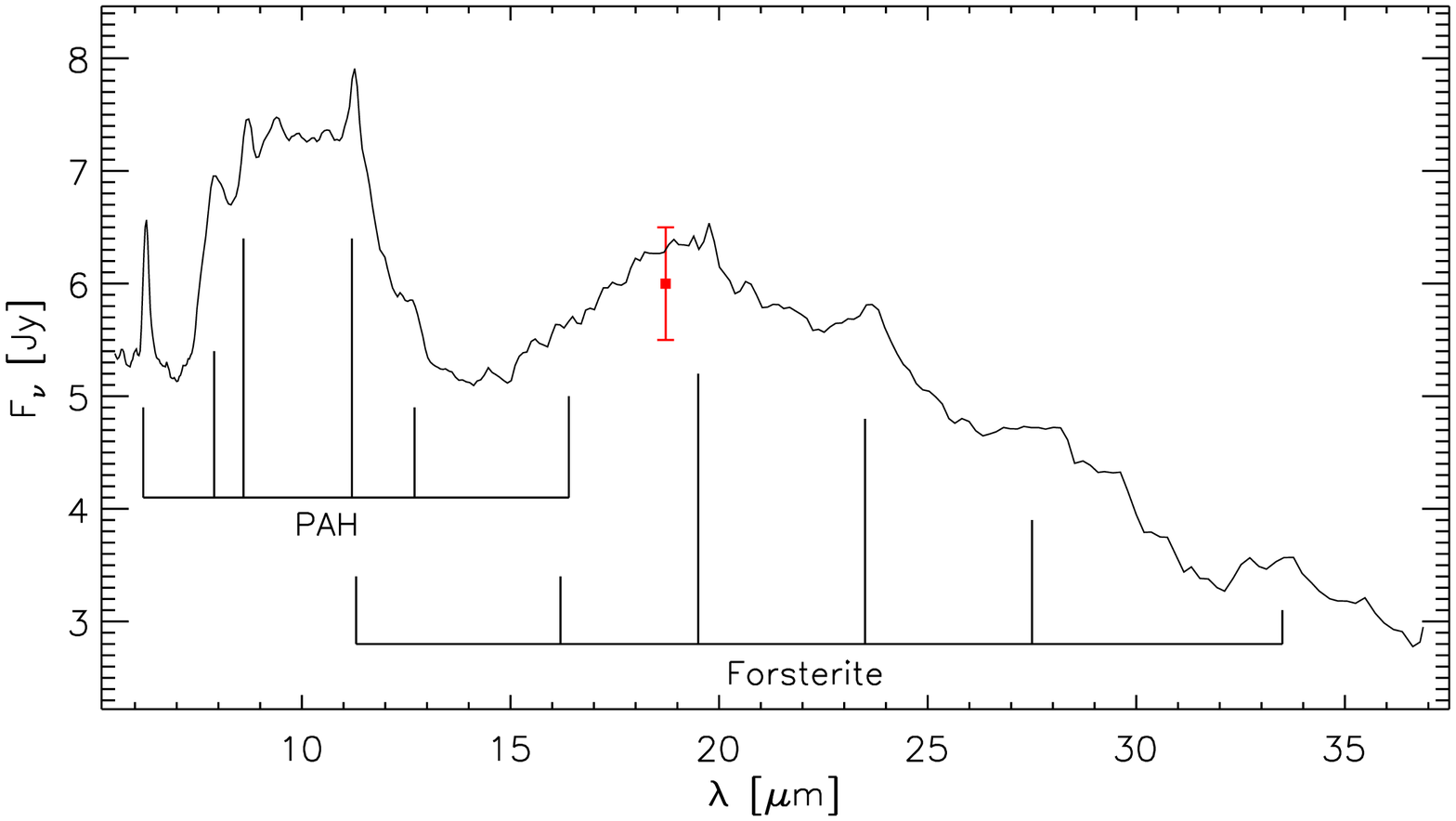}
\caption{The Spitzer LR spectrum of HD\,95881 (SNR $\approx$ 500). The 
left rake \ch{demonstrates the presence of} the PAH bands at 6.2, 7.9,
8.6, 11.2, 12.7, and 16.4\,$\mu$m. The right rake points at the
forsterite bands at 11.3, 16.2, 19.5, 23.5, and 33.5\,$\mu$m. The red
square is the photometry from the VISIR Q-band imaging (see
Sect.\,\ref{sec:ima}).}
\label{fig:spitzer}
\end{figure}
}

\newcommand{\detQ}{
\begin{figure}[t!]
\centering
\includegraphics[width=\columnwidth]{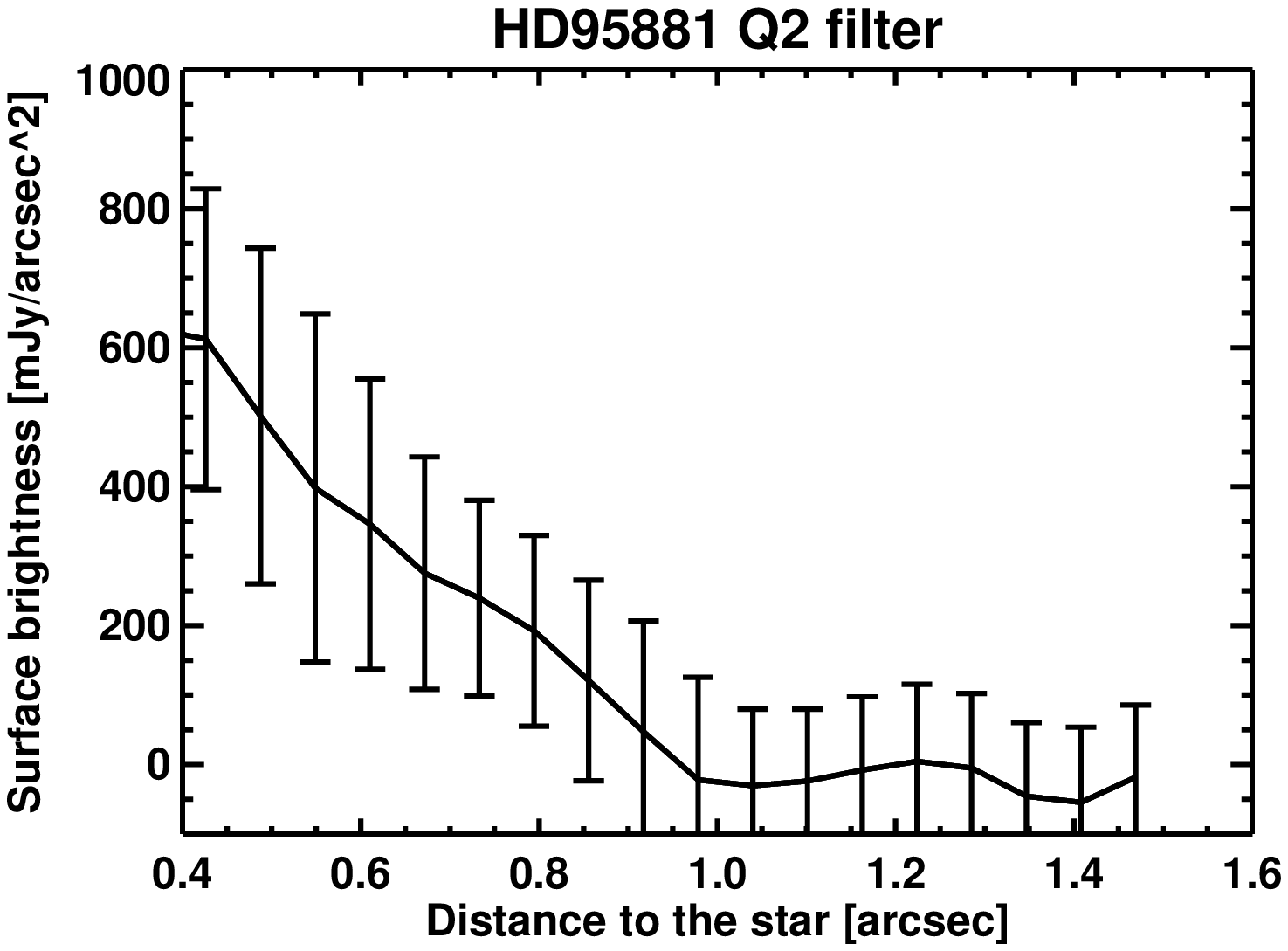}
\caption{Surface brightness levels as a function of distance from the 
star in the Q2 image after point source subtraction. \ch{The error
bars give the 3\,$\sigma$ uncertainty.} }
\label{fig:detection_Q2_ell}
\end{figure}
}

\newcommand{\tenmicron}{
\begin{figure}[t!]
\centering
\includegraphics[width=\columnwidth]{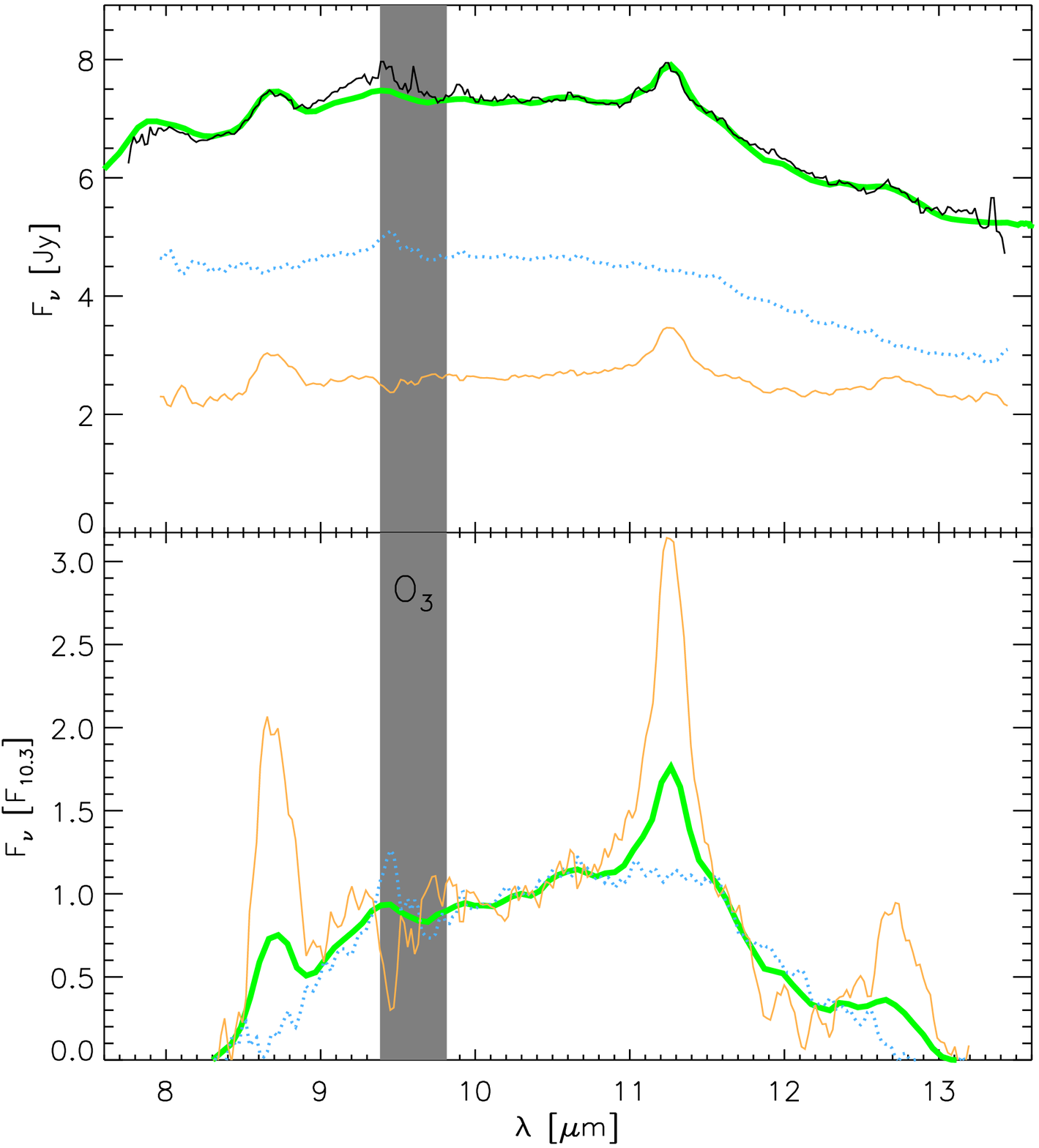}
\caption{\ch{The top panel shows} the VISIR N-band spectrum of 
HD\,95881 (black). Overplotted \ch{are} the Spitzer spectrum
\ch{(thick green)}, the correlated flux as measured with MIDI
\ch{(dotted blue)}, and the difference between the Spitzer and MIDI
spectra \ch{(orange)}. The Spitzer spectrum represents the entire
disk, the MIDI correlated flux spectrum represents the inner disk ($r
\lesssim 2$\,AU), and the difference spectrum represents the outer
disk \ch{($2 \lesssim r \lesssim 100$\,AU)}. \ch{The bottom panel shows} the normalized
continuum subtracted silicate feature (same color coding).  The
similar shape of the inner and outer disk silicate feature indicates a
similar \ch{composition and grain size distribution}.  }
\label{fig:tenmicron}
\end{figure}
}

\newcommand{\Models}{
\begin{figure*}[t!]
\centering
\includegraphics[width=0.98\textwidth]{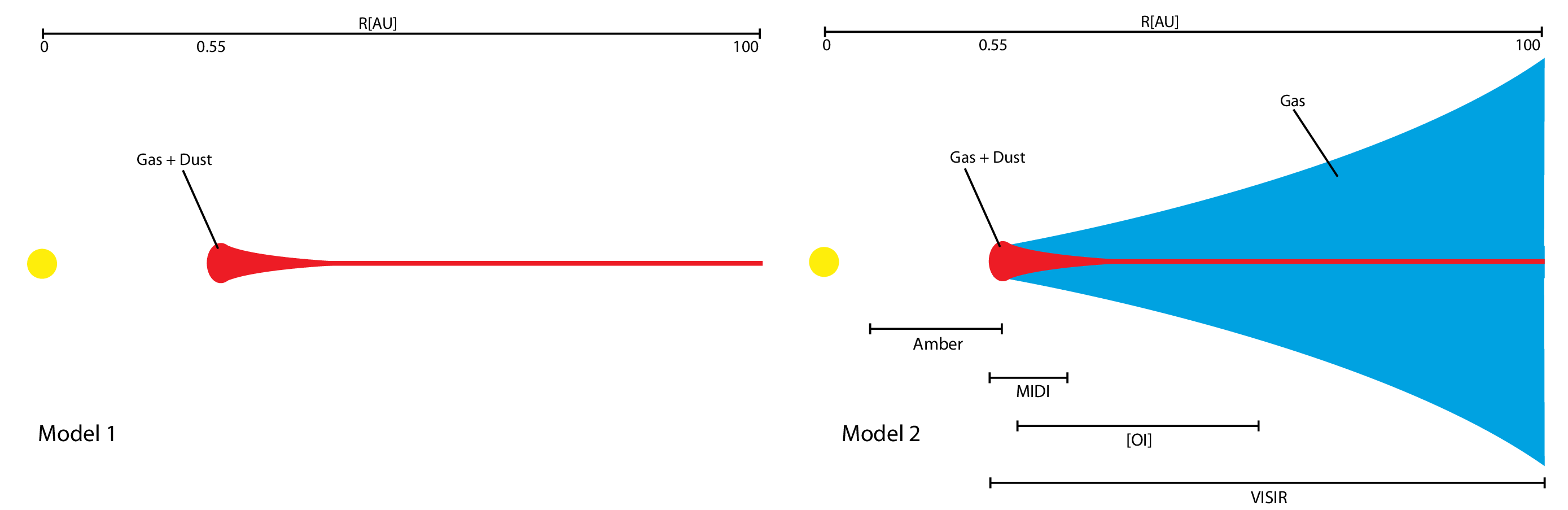}
\caption{Schematic depiction of the two considered disk models. First
we consider a spatial distribution that is the same for both the dust
and the gas. Second we consider a model that assumes that the surface
density of the gas decays much slower with distance from the
star. \ch{In this second model we indicated the different regions of
the disk that are probed by the observations discussed in this
paper.}}
\label{fig:models}
\end{figure*}
}

\newcommand{\modelfitplots}{
\begin{figure*}[t!]
\centering
\resizebox{0.92\hsize}{!}{
\includegraphics{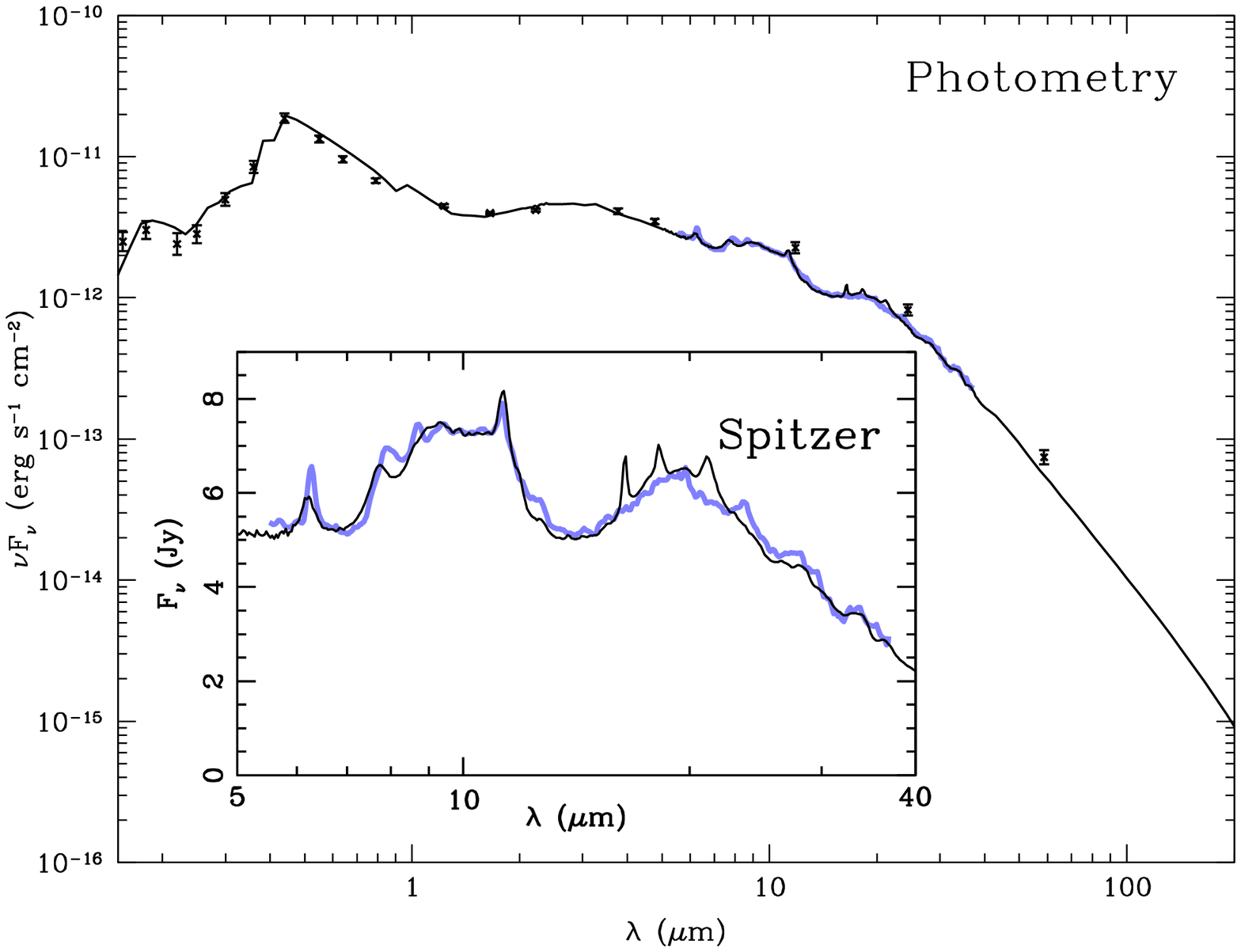}
\includegraphics{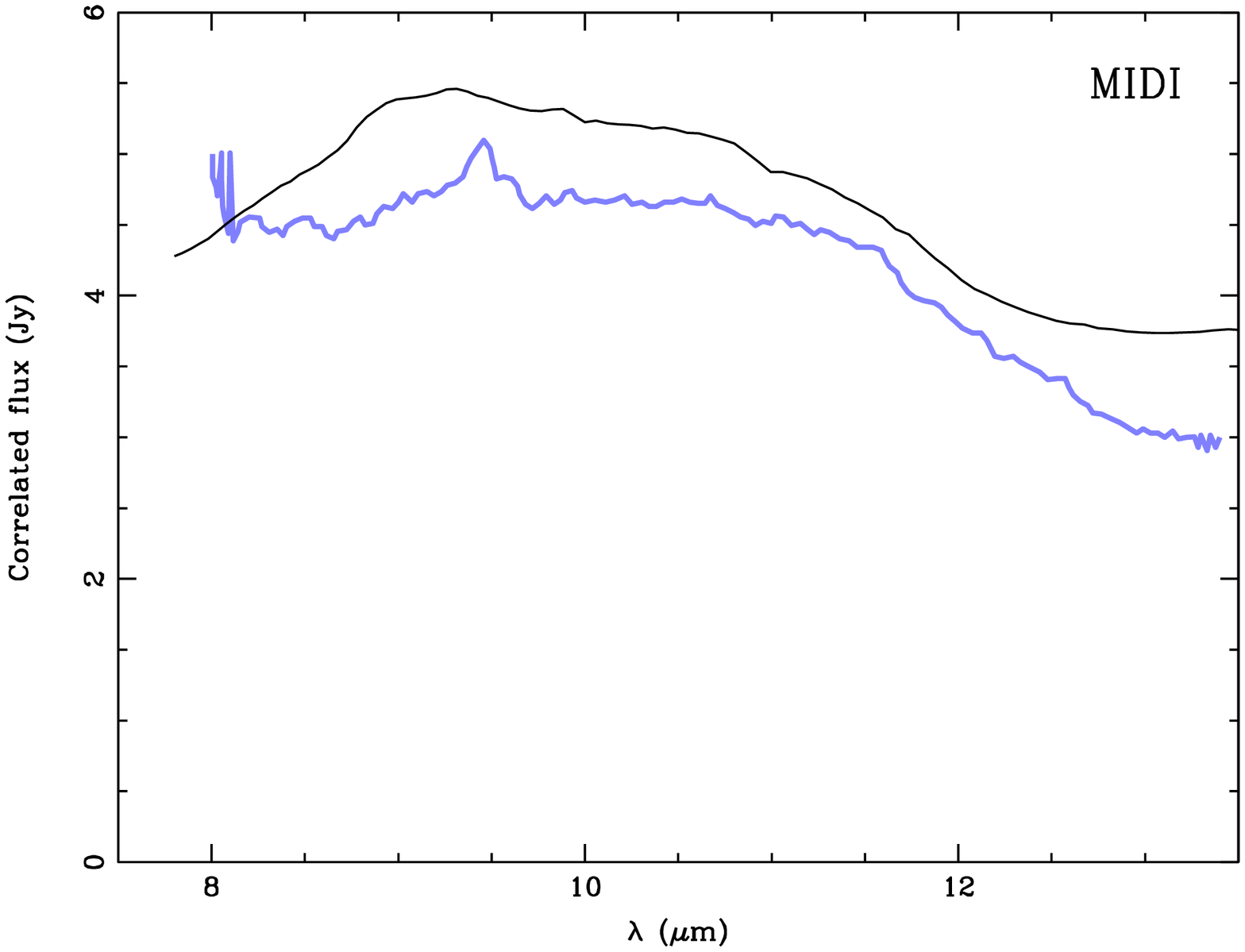}}
\resizebox{0.92\hsize}{!}{
\includegraphics{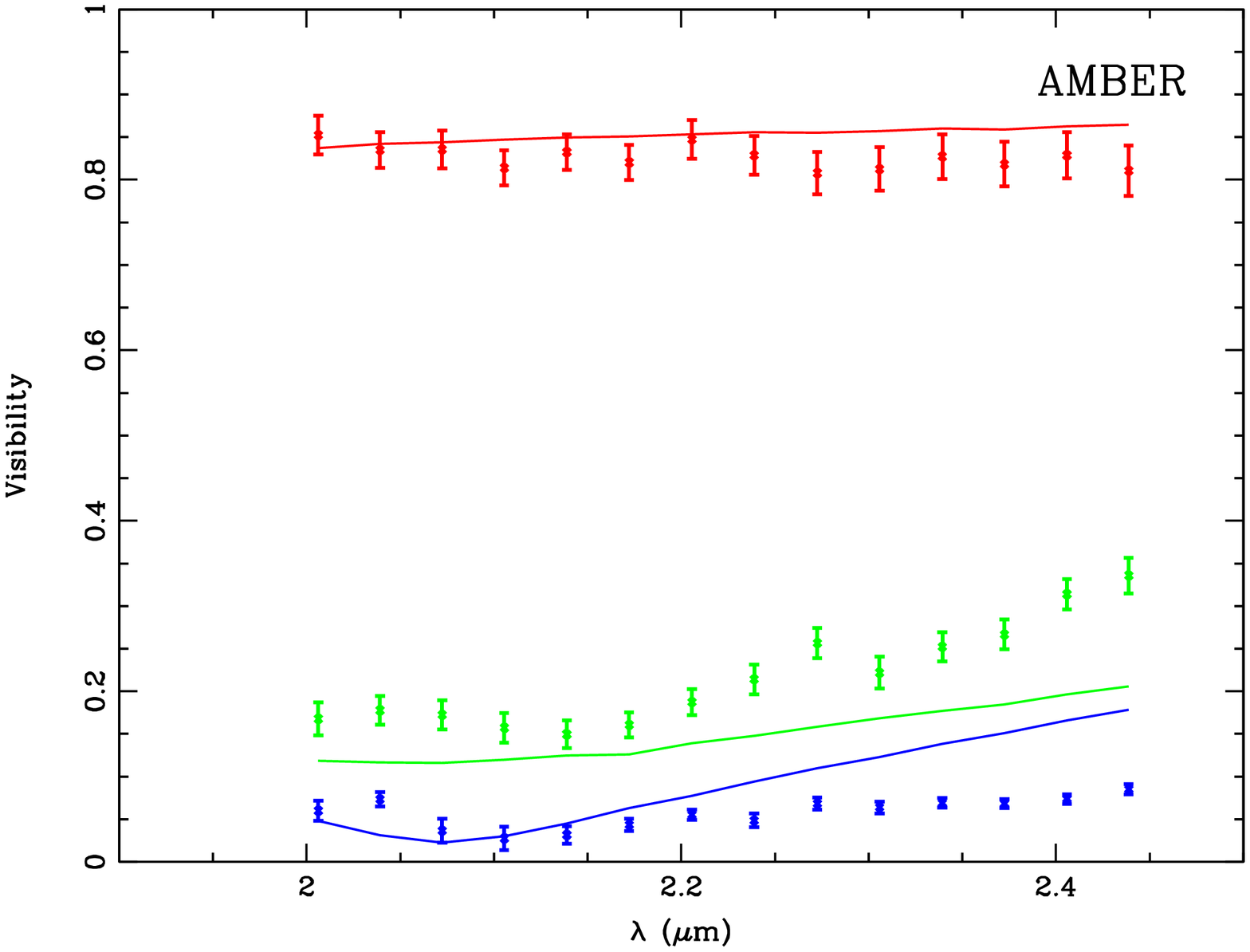}
\includegraphics{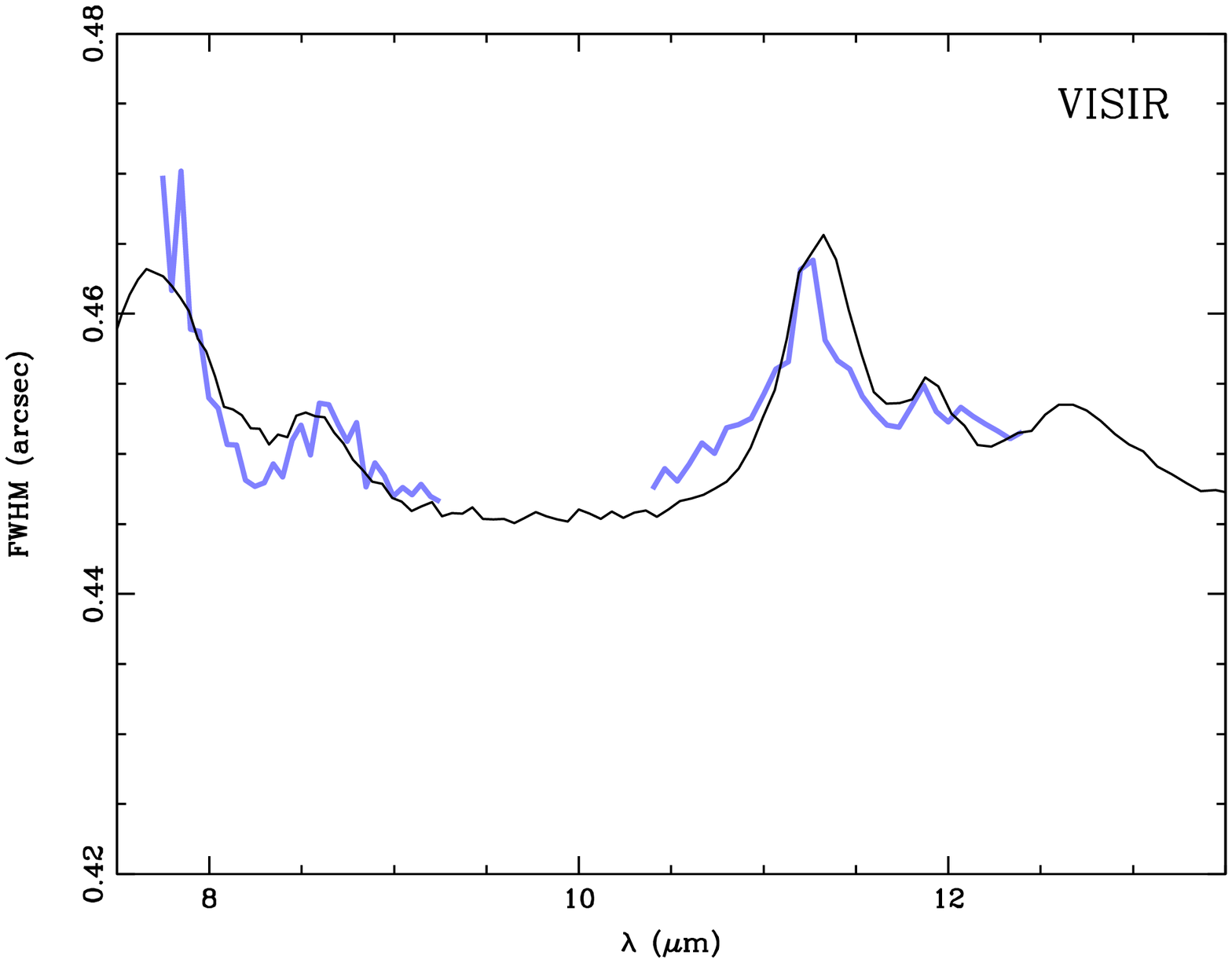}}
\caption{The best model fit to the observables of HD\,95881. The 
upper left panel shows the fit to the spectral energy distribution
with an inset for the Spitzer spectrum. The \ch{blue} line is the
observed spectrum, the black line the model spectrum and the points
with error bars are photometric measurements obtained from the
literature. The upper right panel shows the correlated flux as
obtained by MIDI. The \ch{blue} line is the observed correlated flux
and the black line the model. The bottom left panel shows the
visibilities obtained by AMBER. The colored lines give the model
results (color coding the same as Fig.\,\ref{fig:amber}). The bottom
right panel shows \ch{in blue} the VISIR FWHM as a function of
wavelength and in black the model. }
\label{fig:modelfit}
\end{figure*}
}

\newcommand{\ch}[1]{#1}
\begin{document}
\title{
HD\,95881: A gas rich to gas poor transition disk?
\thanks{Based on observations collected at the European Southern Observatory, Chile.
Under program IDs: 060.A9234A, 076.C-0159A, 077.C-0367A, 078.C-0281}
}

\author{
A.P. Verhoeff \inst{1}
\and M. Min \inst{2}
\and B. Acke \inst{3}\fnmsep\thanks{Postdoctoral Fellow of the Fund for Scientific Research, Flanders.}
\and R. van Boekel \inst{4}
\and E. Pantin \inst{5}
\and L.B.F.M. Waters \inst{1,3}
\and A.G.G.M. Tielens \inst{6}
\and M.E. van den Ancker \inst{7}
\and G.D. Mulders \inst{1}
\and A. de Koter \inst{1,2}
\and J. Bouwman \inst{4}
}

\offprints{A.P. Verhoeff, \email{verhoeff@uva.nl}}

\institute{
Astronomical Institute ``Anton Pannekoek'', 
University of Amsterdam, P.O. Box 94249, 1090 GE Amsterdam, 
The Netherlands
\and Astronomical Institute Utrecht, University of Utrecht, P.O. Box 80000, 
3508 TA Utrecht, The Netherlands
\and Institute for Astronomy, KU Leuven
Celestijnenlaan 200B, B-3001 Leuven, Belgium 
\and Max Planck Institut f\"ur Astronomie,
K\"onigstuhl 17, 69117 Heidelberg, Germany 
\and CEA/DSM/DAPNIA/Service d'Astrophysique, CE Saclay F-91191 Gif-sur-Yvette, France
\and Leiden Observatory, Niels Bohrweg 2, 2300 RA Leiden, The Netherlands
\and European Southern Observatory, Karl-Schwarzschild-Strasse 2, 
85748 Garching bei M\"unchen, Germany
}

\date{Received ....; accepted .....}

\abstract
{Based on the far infrared excess the Herbig class of stars is divided
into a group with flaring circumstellar disks (group I) and a group
with flat circumstellar disks (group II). Dust sedimentation is
generally proposed as an evolution mechanism to transform flaring
disks into flat disks. Theory predicts that during this process the
disks preserve their gas content, \ch{however} observations of
group II Herbig Ae stars demonstrate a lack of gas.}
{We map the spatial distribution of the gas and dust around the group
II Herbig Ae star \object{HD\,95881}.}
{We analyze optical photometry, Q-band imaging, infrared spectroscopy,
and K and N-band interferometric spectroscopy. We use a Monte Carlo
radiative transfer code to create a model for the density and
temperature structure which quite accurately reproduces all the
observables. }
{We derive a consistent picture in which the disk consists of a thick
puffed up inner rim and an outer region which has a flaring gas surface
and is relatively void of \ch{'visible'} dust grains. }
{HD\,95881 is in a transition phase from a gas rich flaring
disk to a gas poor self-shadowed disk.}


\maketitle


\section{Introduction}
Herbig Ae (HAe) stars are known to have gas-rich, dusty disks that are
the remnant of the star formation process. These disks are most likely
the sites of ongoing planet formation. The processes leading to and
associated with planet formation modify both the composition and the
geometry of the disk. Grain growth and grain settling are expected to
result in large spatial variation of the grain size distribution and
the gas to dust mass ratio within the disk. The gravitational
interaction of proto-planets with the disk can create gaps/holes.
Also the gas and the dust chemistry is expected to vary spatially.  In
order to understand planet formation, it is thus important to
establish the spatial distribution of gas and dust in protoplanetary
disks independently.

Observationally, the Spectral Energy Distributions (SEDs) of HAe stars
have been divided into two groups, that reflect differences in the
slope of the mid-IR (10-60 $\mu$m) spectral range
\citep{2001A&A...365..476M}. Group\,I sources have red SEDs, while
group II sources have blue SEDs. These differences can be interpreted
in terms of the geometry of the disk. The direct irradiation of the
inner rim of a disk with an inner hole causes it to be puffed up
\citep{2001ApJ...560..957D}. This puffed up inner rim casts a shadow,
and only the outer disk surface regions emerge from the shadow and
receive direct stellar light. Depending on the dust opacity, some
disks may never emerge from the shadow of the inner rim
\citep{2004A&A...417..159D}. This provides an elegant explanation for
the observed two types of SEDs: group I sources being \emph{flaring},
and group II sources \emph{self-shadowed}.  This interpretation has
been confirmed using spatially resolved mid-infrared (IR) imaging with
the {\em Very Large Telescope Interferometer} (VLTI;
e.g. \citealt{2004A&A...423..537L}).

A difference between group I and group II sources was also found for
the strength of the mid-IR emission bands attributed to Polycyclic
Aromatic Hydrocarbons (PAHs; \citealt{2001A&A...365..476M};
\citealt{2004A&A...426..151A}): flaring disks tend to show strong PAH
emission while self-shadowed sources show weaker or no PAH emission
\ch{(\citealt{2010ApJ...PAHs})}. A similar difference was found for
the strength of the [O\,I]~6300\,\r{A} line
\citep{2005A&A...436..209A}. However, there is significant scatter in
these trends (see below). Both the PAHs and the [O\,I] line strength
probe the \emph{gas} in the upper disk layers, and both require direct
irradiation of the disk surface by stellar photons to be excited. PAHs
mainly probe the disk on scales of several tens to 100 AU
(e.g. \citealt{2004A&A...418..177V}; \citealt{2006Sci...314..621L};
\citealt{2007A&A...476..279G}; \ch{\citealt{2010A&A...HAEBE}}),
i.e. similar scales as the dust continuum emission in the 10-60 $\mu$m
wavelength range. These observations suggest that the spatial
distribution of gas in group II sources is different from that of
group I sources: apparently, in group II sources the gas in the
surface of the outer disk does not receive direct stellar photons.

In a theoretical study, \citet{2007A&A...473..457D} show that for
disks in which the dust settles but the scale-height of the gas does
not change, both group I and group II sources should show prominent
PAH emission, contrary to the observed trend. However, some disks
classified as group II sources (i.e. with a self-shadowed dust
geometry) are observed to show prominent PAH emission and [O\,I] line
emission; examples are HD\,98922 and HD\,95881
\citep{2005A&A...436..209A,
2004A&A...426..151A}. \ch{\cite{2010A&A...HAEBE}} show that for
HD\,95881 the PAH emission is extended at a scale similar to those of
group I sources. \citet{2008A&A...491..809F} \ch{and
\cite{2008A&A...485..487V} studied three HAeBe stars and found that
the gas and dust in their disks has a different spatial distribution.}
In the case of the group II source HD\,101412, PAH and [O\,I] emission
were detected and found to be more extended than the dust continuum at
10\,$\mu$m. These observations suggest that disks exist in which
\emph{the dust has settled but the scale-height of the gas is still
large enough at several tens of AU distance from the star to produce
substantial PAH and [O\,I] line emission}. Such disks may provide
important clues as to how gas-rich disks evolve to gas-poor debris
disks.

In this paper, we study the spatial distribution of the gas and dust
in the disk of HD\,95881. This star was part of a larger study of
spatially resolved mid-IR spectroscopy of HAe stars
(\ch{\citealt{2010A&A...HAEBE}}). We use optical spectroscopy of the
[O\,I] line, the SED, infrared spectra as well as near-IR and mid-IR
interferometric observations to constrain the geometry of the gas and
dust in the disk. We use a hydrostatic equilibrium disk model to fit
the SED and compare the predicted spatial distribution of the near-IR
and mid-IR emission of the best fitting disk model to the
interferometric observations. We find convincing evidence that the
dust in the disk of HD\,95881 has settled but that the gas still has a
significant scale-height. We derive an estimate on the total disk mass
by fitting the strength of the PAH bands.

\parameters
\section{Stellar parameters}
There is little information \ch{available} on HD\,95881 in the
literature. From optical photometry the spectral type was determined
to be A2III/IVe (\citealt{1975mctd.book.....H}), which was translated
into an effective temperature of 8990\,K
(\citealt{2004A&A...426..151A}). The distance was established by a
relatively uncertain association with the star formation region Sco
OB2-4 (\citealt{2004A&A...426..151A}). The distance to this region was
previously determined by \cite{1999AJ....117..354D}. When we take
these parameters to pinpoint HD\,95881 in the
Hertzsprung-Russell-diagram, we find that it is situated to the left
of the Zero Age Main Sequence (ZAMS), which is unphysical. Since we
consider the determination of $T_{\rm eff}$ as reliable, we adopted a
luminosity of 15.4\,L$_{\odot}$ assuming the ZAMS luminosity from
\cite{1994A&AS..103...97M} at the given $T_{\rm eff}$.  The
corresponding ZAMS mass was adopted as the stellar mass. From the new
luminosity, new estimates for the radius and distance follow. Table
\ref{tab:parameters} lists the basic characteristics and our new
estimates. To double check the new distance we looked at the Tycho
parallax, which gave a lower limit of 80\,pc. We also consulted the
Hipparcos catalog for spectral types and $(B-V)$ photometry of stars
within 1$^\circ$ from HD\,95881 and compared the distance to the color
excess $E(B-V)$ for the region (\citealt{1999AJ....117..354D}).  For
HD\,95881 the color excess of $E(B-V) = 0.08$ leads to an upper limit
on the distance of $\sim$200\,pc. The derived distance interval of
80-200\,pc is consistent with both distance estimates, but because of
the argument given above we adopt a distance estimate as derived from
the ZAMS of 170$\pm$30\,pc.

\section{Observations}
\subsection{AMBER}
Spectrally dispersed K-band interferometric observations of HD\,95881
were obtained with VLTI/AMBER on the UT1-UT3-UT4 baseline setting
during the night of February 4, 2007. The weather conditions were
excellent, with the optical seeing as low as 0.5$\arcsec$.  The data
have been reduced according to the standard AMBER data reduction
procedure described in \cite{2007A&A...464...29T}.  AMBER observations
of a standard star (HD\,100901, K0/K1III) were performed directly before
the science measurement and used to calibrate the raw visibilities.

\subsection{VISIR imaging}
Q-band imaging data were obtained on the 14th of December 2005 using
the {\em VLT Imager and Spectrometer for mid-Infrared} (VISIR; see
\citealt{2004Msngr.117...12L}). Standard "chopping and nodding"
mid-infrared observational technique was used to suppress the
background dominating at these wavelength. The Q2 filter was chosen,
which has a central wavelength of 18.72\,$\mu$m and a half-band width
of 0.88\,$\mu$m. The pixel field of view was 0.075$\arcsec$ and the
orientation was standard (north up, east to the left). The total
integration time spend on HD\,95881 was 680\,s. The reference star,
HD\,1024601 (10.6\,Jy in the Q2 filter), was chosen from the database
of VISIR standard stars based on criteria of flux and distance on the
sky to the target. 456\,s of integration time was spent on
HD\,102461. The airmass of both sources was always below 1.5. The
sensitivity achieved was estimated to be
57\,mJy\,/\,10\,$\sigma$\ch{/}1h\ch{;}. The optical seeing was
moderately fluctuating in the range 0.75-0.85$\arcsec$,
\ch{corresponding to an estimated Q-band seeing of
$\lesssim$\,0.4$\arcsec$.}

\subsection{VISIR spectroscopy}
Long slit N-band spectra were obtained with VISIR in the low
resolution (LR) mode. A sample of 17 HAe stars was observed under the
VISIR GTO program on circumstellar disks
(\ch{\citealt{2010A&A...HAEBE}}). During the nights of December 16,
2005 and April 8, 2006 we observed HD\,95881. Standard chopping and
nodding \ch{along the slit} was used with a chopper throw of
8$\arcsec$, a slit-width of 0.75$\arcsec$, and a standard North-South
orientation. After and before the science measurements bright standard
stars were observed. The airmass of the observations was $\sim$1.6 and
the optical seeing was around 1.0$\arcsec$ during the first night and
around 1.2$\arcsec$ during the second night, \ch{corresponding to an
estimated N-band seeing of $\lesssim$\,0.65$\arcsec$.}

\subsection{MIDI}
HD\,95881 was observed with the {\em MID-infrared Interferometric
instrument} (MIDI; see \citealt{2003SPIE.4838..893L}), the 10\,$\mu$m
instrument of the VLTI, during the night starting June 6, 2004, as
part of the "science demonstration" program. The UT1-UT3 baseline was
used, resulting in a projected baseline length of 65\,m at a position
angle of 81$^{\circ}$ east of north. The grism was used to spectrally
disperse the signal at a resolution of R = $\lambda$/$\Delta\lambda$ =
230. This ensures that relatively narrow "dusty" emission features
such as those of crystalline silicates and PAHs are well resolved
spectrally. The seeing was constant at a value of 0.6$\arcsec$ and the
atmospheric transparency was excellent.

We performed observations in "High-Sens" mode, i.e. we took an
interferometric measurement combining the light from both telescopes,
and subsequent photometric measurements \ch{of} the signal from one
telescope at a time. Stars of known brightness and angular diameter
were observed, using the same procedure, for photometric calibration
and monitoring system coherence losses (interferometric "transfer
function"). We found our interferometric measurement of HD\,95881 to
be of significantly higher quality than the corresponding photometric
measurements, and chose to directly calibrate the correlated flux
rather than converting to interferometric visibility by division over
the photometry.

\subsection{Additional data}
\label{sec:add}
A low resolution Spitzer-IRS spectrum is used to compare and
flux-calibrate the VISIR spectrum. For the acquisition and reduction
of the Spitzer data we refer to \ch{\cite{2010ApJ...Spitzer}}.
Photometric data points were taken from
\cite{2004A&A...426..151A}. Together with the Spitzer spectrum they
make up the SED, which allows us to constrain the disk geometry (see
Sect.\,\ref{sec:SED}). Finally, [O\,I] data were taken from an earlier
study, see \cite{2005A&A...436..209A}.

\section{Analysis}
\amber
\subsection{AMBER}
\label{sec:amber}
A good fit (\,$\chi^2$/$\nu$\,=1.8) to the AMBER data could be
obtained with an inclined uniformly emitting ring surrounding a
point-like source representing the star (Fig.\,\ref{fig:amber}). The
inner and outer radius of this ring are 2.2$\pm$0.2\,mas (0.37\,AU)
and 2.7$\pm$0.3\,mas (0.46\,AU).  The disk inclination is
60$^\circ\pm$10$^\circ$ and its position angle is
102$^\circ\pm$2$^\circ$. The mean radius of the ring is 2.4\,mas
($\sim$0.4\,AU). This is similar to the values found by
\cite{2004ApJ...613.1049E} for a group of Herbig Ae/Be stars. The
fraction of the K-band flux that is ascribed to the star by the fit of
a Kurucz model to the optical photometry ($\sim$20\%) is not exactly
equal to the AMBER estimate of 32$\pm$4\%. It may be that this
discrepancy is due to the presence of an unresolved disk component,
e.g. hot gas \ch{well within} the dust sublimation radius, that
contributes to the K-band emission. \ch{Gas emission from a very
compact region around the star, where no dust species can survive, has
been claimed for quite a number of pre-main-sequence stars}
(\citealt{2007ApJ...657..347E}; \citealt{2008ApJ...676..490K};
\citealt{2008A&A...485..209A}; \citealt{2008ApJ...677L..51T}; and
\citealt{2008A&A...483L..13I}). \ch{However, refractory dust has also
been suggested to be present in the innermost regions
(\citealt{2009arXiv0911.4363B}).}

\profile
\subsection{[O\,I] data}
\cite{2005A&A...436..209A} have investigated the [O\,I]~6300\,\r{A}
emission line in a large sample of Herbig stars. The authors argue
that the emission is non-thermal and originates from the disk surface
of a flared circumstellar disk. This is consistent with the large
detection rate of [O\,I] emitters among the group I sources in the
sample. Roughly half of the group II sources, however, display the
[O\,I]~6300\,\r{A} line in emission as well, albeit less
strong. HD\,95881 is one of these targets.

In Fig.\,\ref{fig:profile}, the [O\,I]~6300\,\r{A} line profile is
shown. We have fitted a generic model to the data, assuming that the
intensity drops off with radius as a power law and that the disk is in
Keplerian rotation. We adopt the disk inclination derived from the
AMBER fit. An acceptable fit \ch{(reduced $\chi^2$ = 3}) was achieved
with a power index of $-$3.0$\pm$0.2, an inner radius of
0.9$\pm$0.2\,AU and an outer radius larger than 11\,AU. The best fit
\ch{(reduced $\chi^2$ \ch{= 1.5})} has an outer radius beyond
31\,AU. In the figure we show the best-fit profile, as well as the
profile that corresponds to the model with an 11\,AU outer
radius. Decreasing the outer radius of the model to even lower values
would further deteriorate the fit quality. We can therefore conclude
that a significant fraction of the [O\,I] emission comes from a region
at tens of AU from the star. Note that our model does not account for
the 15\% of [O\,I] flux which is emitted in the blue- and red wings of
the profile. In these regions, close to the star, the power law
approximation fails.

\spitzer
\subsection{Spitzer}
\label{sec:spitzer}
The Spitzer-IRS spectrum is given in Fig.\,\ref{fig:spitzer}. It
displays a rich mineralogy, there is emission of amorphous and
crystalline silicates as well as emission of various PAH
bands. \ch{See \cite{2010ApJ...Spitzer} for a detailed discussion of
the mineralogy.} The blue slope at longer wavelengths is typical for
group II sources.

Compared to other HAe stars HD\,95881 appears \ch{to be a} very
ordinary \ch{PAH emitter}. \ch{We have verified this by comparing PAH
band-strength ratios.} We have taken the 6.2, 7.7, 8.6, 11.2 and
12.7\,$\mu$m PAH band strengths from \ch{\cite{2010A&A...HAEBE}} and
\cite{2010ApJ...PAHs} and looked at the ratios of their continuum
subtracted and integrated strengths. \ch{The only aspect that makes
HD\,95881 stand out slightly} is the 8.6\,$\mu$m PAH feature, which is
relatively strong.

\subsection{VISIR imaging}
\label{sec:ima}
A dedicated data reduction pipeline was used for the imaging. It
features a comprehensive set of methods to correct for instrumental
signatures such as detector striping or background low-frequencies
excess of noise (\citealt{2008SPIE.7014E..70P,2009svlt.conf..261P}).
A photometric analysis gives an integrated flux for HD\,95881 of
6.0$\pm$0.5\,Jy in the Q2 filter (18.72\,$\mu$m). This is in fair
agreement with the Spitzer data (see Fig.\,\ref{fig:spitzer}). Since
the emission of the star is negligible at this wavelength, this flux
can be attributed to the disk. We searched for an extended emission
component using PSF subtraction at a sub-pixel (1/10) precision
level. The PSF was derived from the observation of the standard
star. The resulting residuals display an excess of signal that is
roughly axi-symmetric and decreases gradually as a function of
distance from the star.

Assuming that the disk has the geometric parameters given in
table\,\ref{tab:modelfit}, we divided the residuals in a series of
concentric ellipses having a separation of 0.075$\arcsec$ along the
semi-major axis of the disk. We assumed an aspect ratio corresponding
to a flat disk inclined at 55$^\circ$, and a position angle of
103$^\circ$ from North. We used these elliptic annuli to numerically
estimate the surface brightness distribution and the corresponding
uncertainty levels.

\detQ Our statistical analysis confirms we have a \ch{significant}
detection since for distances in the range 0.4-0.85$\arcsec$ from the
star the average values in each of the elliptical annuli are well
above the detection limits set at 99\% confidence. The measured
average surface brightness in elliptic annuli as a function of the
semi-major axis distance to the star are plotted in
Fig.\,\ref{fig:detection_Q2_ell}. The 3\,$\sigma$ uncertainties are
displayed as bars on the plot. The total flux in the resolved
component is 0.7$\pm$0.1\,Jy.

\tenmicron
\subsection{VISIR spectroscopy}
\label{sec:spc}
The general reduction and analysis strategy of the spectroscopic VISIR
data is described in detail in \ch{\cite{2010A&A...HAEBE}}. For
HD\,95881 the telluric correction was done by means of an airmass
interpolation of two calibrators. The observation of April 8, 2006,
appeared to be suffering strong atmospheric residuals. The spectrum of
December 16 2005 was thus chosen as most representative. The missing
observation of the 9.8\,$\mu$m setting was replaced with the poor one
from April 8, 2006. We scaled the VISIR spectrum to the Spitzer
spectrum at 10.6\,$\mu$m using \ch{an average factor for the different
settings of 1.16}. The resulting spectrum in Fig.\,\ref{fig:tenmicron}
has a SNR of $\sim$300. The agreement with the Spitzer spectrum is
encouraging. The slight deviation observable just left of the ozone
band at 9.6\,$\mu$m is typical for the quality of the data taken on
April 8, 2006.

The Full Width at Half Maximum (FWHM) of the spatial emission profile
of the target was determined by performing a Gauss-fit in 32 merged
wavelength bins. Comparison of the science signals with the PSF shows
that the target is unresolved in the continuum. After quadratic
subtraction of the PSF and averaging over the median values of all
measurements we find a three sigma upper-limit to the FWHM extent of
the continuum emission region of $<$\,0.46$\arcsec$, which corresponds
to $<$\,79\,AU at the adopted distance of 170 pc. However the science
signal displays a relative increase at 8.6 and 11.2\,$\mu$m and an
upturn to the left of $\sim$8\,$\mu$m, which are exactly the
wavelengths at which the PAH molecules have emission features \ch{(see
\ch{\citealt{2010A&A...HAEBE})}}. We checked the significance of these
FWHM features with respect to pixel-to-pixel variations and concluded
that {\em the PAH emission is significantly more extended than the
continuum}.

In order to estimate the spatial extent of the PAH emission we
measured the spatial emission profile at the peak wavelengths of the
PAH bands and we subtracted the spatial emission profile of the
continuum contribution. This continuum profile was determined by
interpolating the intensities and spatial profiles adjacent to the PAH
bands. The resulting observed spatial profile of the PAH emission was
Gaussian fitted to obtain the FWHM. Finally, the instrumental width
(i.e. the PSF) was quadratically subtracted to obtain a measure for
the intrinsic extent of the PAH emission. We found FWHM values of
0.34$\arcsec^{+0.05}_{-0.08}$ and 0.39$\arcsec^{+0.04}_{-0.06}$ for
the 8.6 and 11.2\,$\mu$m PAH bands respectively, which results in
absolute sizes of 58 and 66\,AU. In a Gaussian distribution of the PAH
emission this would mean that 99\% is confined in a radius of
$\sim$100\,AU. Note that this is a conservative estimate of the PAH
emission scale since the PAH surface brightness is expected to fall
off with distance from the star together with the \ch{stellar} flux as
1/$r^2$.

\subsection{MIDI}
In Fig.\,\ref{fig:tenmicron} we compare the spectrum \emph{in
correlated flux} as seen by MIDI, \ch{i.e. the total flux times the
visibility,} to the spectra observed by VISIR and Spitzer. The
correlated flux spectrum is dominated by the central few AU of the
disk, the Spitzer and VISIR spectra probe the entire disk. Note the
difference in the strength of the 8.6 and 11.2\,$\mu$m PAH bands in
the spectra. These bands are prominent in the total flux spectra, but
essentially absent in the correlated flux spectrum. The PAH emission
region is apparently outside of the disk region probed by MIDI. This
shows by direct measurement that {\em the PAH features arise at scales
much larger than $\sim$2\,AU in the disk of HD\,95881}. To stress this
point we also plotted the difference between the Spitzer and the MIDI
correlated flux spectrum in \ch{the bottom panel of}
Fig.\,\ref{fig:tenmicron}. This difference spectrum is dominated by
the emission of the outer disk \ch{($2 \lesssim r \lesssim 100$\,AU)}
and shows very distinct PAH features.

To investigate spatial differences in the weak silicate emission we
considered the shape of the feature in the Spitzer, MIDI, and
difference spectrum in a consistent manner. We approximated a
continuum with a straight line intersecting the spectra at 8.3 and
13.2\,$\mu$m, subtracted this from the spectra and then normalized the
spectra with the flux level at 10.3\,$\mu$m. The bottom panel of
Fig.\,\ref{fig:tenmicron} shows the result. The shape of the silicate
feature \ch{inside and outside a radius of $\sim 2$\,AU} is very
similar, which implies that the \ch{silicate composition and grain
size distribution} should also be very similar. \ch{This is not
surprising, since the high visibility indicates that the silicate
emission is dominated by a compact inner region ($r \lesssim
10$\,AU).}

\subsection{Observational picture}
Before we describe a detailed modeling effort of the circumstellar
material, we summarize the analyses of the various observations. This
already gives an insight into the spatial distribution of the gas and
the dust. \ch{In Fig.\,\ref{fig:models} we display schematic
representations of two disk models, that will be considered in
Sect.\,\ref{sec:SED}. In the second model we indicated the diagnostic
scales of the various data sets to combine them into a consistent
picture.}

The AMBER data probes the very inner parts of the disk. The analysis
shows that the K-band emission could be explained with an emitting
ring at $\sim$0.4\,AU. A more physical model, which will be presented
in Sect.\,\ref{sec:SED} has the inner rim of the disk at 0.55\,AU. A
part of the K-band emission was shown to \ch{come from close to the
star, and well within the inner dust disk (represented by the ring).}
This could be an indicator of ongoing accretion.

The [O\,I]~6300\,\r{A} emission line is formed by the
photo-dissociation of OH molecules by UV photons
\ch{(\citealt{2005A&A...436..209A})}. The detection of the
[O\,I]~6300\,\r{A} at large distances (from one to tens of AU) from
the central star is thus an indication that the outer disk has an
illuminated gas surface.

The Spitzer data establish the presence of PAH emission and the VISIR
spectrum pins it down to a circumstellar disk. The emission features
of PAH molecules are caused by internal vibrational modes, which are
mainly excited by UV photons. Since the PAH molecules are coupled to
the gas, the PAH emission is another indicator of an illuminated gas
surface. The resolved VISIR spectrum sets the radial scale of this gas
surface at $\sim$100\,AU.

The VISIR Q-band image displays a faint extended \ch{axi-symmetric}
emission component ($\sim$10\% of the total flux), that stretches out
to large radii (r $\sim$150\,\ch{AU}).

The MIDI correlated flux spectrum shows that the PAH features
originate from radii much larger than $\sim$2\,AU and that the
silicate composition is quite similar in the inner and outer disk.

\section{Modeling}
\label{sec:SED}
In this section we present a \ch{model} for the disk around HD\,95881
which quite accurately reproduces the observations described above. To
obtain this model we use the Monte Carlo radiative transfer code MCMax
by \cite{2009A&A...497..155M}. This code can compute a self-consistent
disk structure and a full range of observables. It has a build-in
option that models the full PAH excitation, using the temperature
distribution approximation (see e.g. \citealt{1989ApJ...345..230G};
\citealt{1992A&A...266..501S}) including multi-photon events for the
excitation. We use MCMax here to compute the temperature structure,
the vertical density structure and the resulting SED, the Spitzer
spectrum, the AMBER visibilities, the MIDI correlated flux, the VISIR
images, and the FWHM as function of wavelength. The steps to come to
the final model presented here were the following.

\subsection{Initial constraints}
First we fixed the composition, and the size and shape distribution of
the silicate component of the dust to be equal to that obtained from
the 10 micron silicate feature by \citet{2005A&A...437..189V} . For
HD\,95881 these are 80\% large (1.5\,$\mu$m) pyroxene grains, 11\%
large enstatite grains, \ch{1\% small (0.1\,$\mu$m) forsterite
grains,} 5\% large forsterite grains, and 3\% small (0.1\,$\mu$m)
silica grains. \ch{For the shapes of the grains we use the
Distribution of Hollow Spheres \citep[DHS,
see][]{2005A&A...432..909M}, as outlined in
\citet{2005A&A...437..189V}, where we refer to for further details on
the model for the silicate dust grains.} In order to get the required
continuum opacity needed we added amorphous carbon grains. To model
the irregular shape of the carbon grains we \ch{also used DHS}. For
the refractive index of carbon we adopted the data by
\citet{1993A&A...279..577P}. Note that \ch{the} continuum component is
most likely not all in the form of amorphous carbon. Small grains of
metallic iron and/or iron sulfide have extinction properties similar
to that of carbon. Also, large grains of various dust species could
produce the observed continuum component. The abundance of amorphous
carbon is a fitting parameter. \ch{We have to realize that besides
this observable dust component, a 'hidden' component of large grains
probably exists deep inside the disk. Information on this dust
component, which might contain most of the dust mass, can only be
obtained through millimeter observations which are currently not
available for this source. In the following we will use the term
'visible' for the dust grains we can address with the current set of
observations.}

The other free parameters all have to do with the geometry
of the disk. As a first step we focused on the thermal dust grains,
ignoring the PAH bands. The density distribution of the dust disk was
parameterized using a radial surface density
(\citealt{2008ApJ...678.1119H})
\begin{equation}
\Sigma(r) \propto r^{-p}\exp \left\{- \left( \frac{r}{R_0} \right) ^{2-p}\right\},
\end{equation}
for $R_\mathrm{in} < r < R_\mathrm{out}$. Here $R_0$ is the turnover
point from where an exponential decay of the surface density sets in
and $p$ sets the powerlaw in the inner region. We fixed this powerlaw
to $p=1$, a commonly used value (see
e.g. \citealt{2006ApJ...640L..67D}). \ch{Using $p=1$ basically implies
that we assume the viscosity of the disk to linearly increase with
radius. For a discussion on the possible values of $p$ and their
implications for the origin of turbulence and viscosity in the disk we
refer to \citet{2009ApJ...701..260I}.} The vertical density structure
was computed from hydrostatic equilibrium. However, since we found
that this vertical structure cannot reproduce the SED, we included a
scaling parameter $\Psi$ by which the scale-height of the \ch{entire}
disk is increased. For the density distribution we thus have the inner
and outer radii $R_\mathrm{in}$ and $R_\mathrm{out}$, the turnover
radius $R_0$, the total dust mass in \ch{visible} grains
$M_\mathrm{dust}$, and the vertical density scaling parameter $\Psi$,
as free parameters.

\subsection{Fitting the SED}
\ch{In order to get the observed peak over continuum ratio of the 10
and 20 micron silicate features observed in the Spitzer spectrum, we
find that the mass in amorphous carbon grains has to be 25\% of the
total dust mass. Again, we stress that this component represents all
possible sources of continuum opacity, including very large
grains. Second, we constrained the parameters describing the density
structure.} We found that the vertical height of the disk needs to be
significantly increased compared to hydrostatic equilibrium in order
to obtain the large near IR excess \ch{(see also
\citealt{2009A&A...502L..17A} for a discussion on the inner rim
height)}. The height of the disk was scaled with a factor
$\Psi=2.75$. The exponential decay of the surface density sets in at
$R_0=2.5$\,AU. To put the density structure in some perspective, at
$\sim$7.5\,AU our prescription produces the same density as does a
1/$r^2$ density law, but beyond 10\,AU the surface density becomes
negligible. The inner radius of the disk is at 0.55\,AU, while the
outer radius has no influence on the observationally constrained part
of the SED as long as it is beyond $\sim$10\,AU. We find a total mass
in \ch{visible} dust grains of $10^{-8}$\,M$_{\odot}$. Note that the
total dust mass is probably much higher, because the mass in large
dust grains is not constrained.  To this dust distribution we added
PAHs in an abundance needed to explain the features seen in the
Spitzer spectrum. For the opacity of the PAHs we use those computed by
\citet{2001ApJ...551..807D} for molecules consisting of 100 carbon
atoms.

\Models
\subsection{Spatial distribution of the gas}
We consider two possibilities for the spatial distribution of the gas,
as traced by the PAHs (see Fig.\,\ref{fig:models}). The first is to
assume that the gas and the dust have the same spatial distribution in
the disk. The second is to assume that the gas does not have the
exponential decay of the surface density for radii larger than
2.5\,AU, but that this disappearance of the dust at these radii is
caused by grain growth and settling, which do not affect the gas.

The first model, which has the gas and dust in the same spatial
distribution, results in a fairly large abundance of PAH molecules
needed to explain the strength of the features seen. The PAHs are in
this case fairly well shielded from the stellar radiation by the dust
grains, and thus a large amount is needed. Furthermore, in order to
explain the absence of PAH emission in the MIDI correlated flux, we
find that we have to remove the PAHs from the inner 2\,AU. The total
mass in PAHs in this case is 3$\cdot$10$^{-8}$\,M$_{\odot}$. This
model does not reproduce the increase in FWHM at the wavelength of the
PAH features that we found in the VISIR spectroscopic data
(see Sect.\,\ref{sec:spc}).

The second model, which assumes the gas is distributed in a more
extended flared disk (see Fig.\,\ref{fig:models}) resulted in a very
good overall fit of all the observables presented above. In this model
we put the PAHs in a disk with a similar surface density as the dust
grains but with $R_0 = \infty$, i.e. the surface density remains a
powerlaw ($p = 1$) for all radii. In this way we create a flaring
outer gas disk, which is able to catch much of the radiation from the
star. We find that the PAH emission is dominated entirely by the outer
regions. If the PAHs would be destroyed according to the mechanism
proposed by \citet{2007A&A...473..457D} we find that the inner 25\,AU
should be free of PAHs. However, even without destruction the
\ch{contribution to the PAH emission} from these inner regions is
negligible. Thus, we cannot confirm whether or not PAH destruction
takes place in this disk. Extrapolating the powerlaw surface density
distribution of the PAHs to the inner edge (at 0.55\,AU) we find that
at $R_{\rm in}$ the PAH abundance is 0.25\% of the \ch{visible} dust
mass. The total PAH mass we find is 5$\cdot$10$^{-9}$\,M$_{\odot}$,
which is large compared to the mass in \ch{visible} dust grains,
because the PAH disk is so much larger. Including PAH destruction in
the inner disk would only lower the total PAH mass by 12\%. \ch{Note,
that the total dust mass, including the well shielded midplane grains,
can be much higher than the mass in visible grains and the PAH
mass. This could be confirmed by future millimeter observations.}

\modelfit 
\subsection{Final model}
The second model, that assumes that the PAHs do not have an
exponential decay with radius, reproduces the available spatial
information much better. This second model was fine-tuned to best
reproduce all available observables. The position angle and
inclination of the disk were constrained using the interferometric
data. We find that the AMBER interferometric observations put
important constraints on the parameters for the inner regions of the
disk, lifting some of the degeneracies present when these data are not
considered. The parameters of the final model are summarized in
table\,\ref{tab:modelfit}.

\section{Discussion}
\subsection{Comparison of the model with the observations}
Our disk model \ch{accurately reproduces} all the observables. The
fits to the SED, the Spitzer spectrum, the MIDI correlated flux, the
AMBER visibilities, \ch{and the VISIR spectroscopic FWHM are presented
in Fig.\,\ref{fig:modelfit}.}

\modelfitplots 
\emph{Spitzer.} The general slope and most features in the Spitzer
spectrum are reproduced. However, the observed 6.2, 7.9 and 8.6\,$\mu$m
PAH features are stronger than in our model and the predicted PAH
features around 20\,$\mu$m are not seen in the Spitzer spectrum. These
differences are related to the PAH chemistry, which is a much debated
subject \ch{(see e.g. \citealt{2007ApJ...657..810D})}.

\emph{MIDI.} The N-band correlated flux obtained by MIDI has an error
on the absolute calibration of approximately 10\%, similar to the
difference with the model output. Thus our modeling of the spatial
distribution of the \ch{visible} dust grains is consistent with the
MIDI result.

\emph{AMBER.} The simple flat ring+point-source model of
Sect.\,\ref{sec:amber} actually reproduced the inclination, position
angle and visibilities of the disk very well, although the inner rim
radius (0.37\,AU) is significantly below the 0.55\,AU of our final
model. Our more physical final model gives a slightly poorer fit to
the visibilities. This indicates that the structure of the inner disk
is more complicated than assumed. The exact structure is currently
much debated (see Sect.\,\ref{sec:model}).

\emph{VISIR spectrum.} The model FWHM was obtained by making a
Gaussian fit to the spatial profile after convolving the model with a
Gaussian of the same width as the PSF of the VISIR 11.2\,$\mu$m
setting. \ch{The PSF of the 8.5\,$\mu$m setting was scaled to the
model.}  The matching continuum levels are thus a result of our
method, but the increase in the FWHM at the PAH wavelengths are a
confirmation of the correct modeling of the spatial distribution of
the \ch{PAH emission}.

\emph{VISIR image.} A model of the Q2 image was obtained by taking the
output of our model at 19.0\,$\mu$m, just next to the artificial PAH
feature. We convolved this model image with the VISIR PSF and we
applied a PSF subtraction in analogy to the image analysis of
Sect.\,\ref{sec:ima}. The \ch{observed resolved emission component
is not reproduced} by the model. A likely explanation for this is
the photoluminescence of \ch{particles with sizes between dust grains
and PAH molecules, the so called very small grains}, which are not
included in our model (see \citealt{2006A&A...453..969F}).

\subsection{The distribution of gas and dust}
\label{sec:model}
Our modeling reveals that the spatial distribution of the gas and the
dust in the disk around HD\,95881 is different. The dust in the upper
layers of the outer disk is heavily depleted, while the gas still has
a large scale height. To reproduce the near IR flux we needed to
assume a scale height, which is much larger than would be obtained
from vertical hydrostatic equilibrium ($\Psi = 2.75$). We do not have
an explanation for this, but we note that the value of $\Psi$ is
sensitive to the assumptions on the structure and composition of the
material in the inner disk \citep[see e.g.][]{2008A&A...483L..13I,
2008ApJ...689..513T}. Furthermore, our value is well within the range
of scaling parameters (1.0-3.0) needed by \cite{2009A&A...502L..17A}
to explain the near IR SED of about half of their sample of $\sim$30
Herbig Ae/Be stars.

The spatial distribution of the \ch{visible} grain component was well
constrained by the SED. The large grains could not be constrained
directly, but are probably abundantly present in a settled outer
disk. Observations of the millimeter flux will help to constrain the
mass in this component as well as the size distribution. We can
however make an estimate of the total dust mass, based on the modeled
total PAH mass (5$\cdot$10$^{-9}$\,M$_{\odot}$) and the modeled PAH
abundance at the inner edge (0.25\%). \ch{Assuming that at the inner
rim all the dust is visible and} assuming that the PAH to dust ratio
is homogeneous throughout the disk we find that the total dust mass
adds up to 2$\cdot$10$^{-6}$\,M$_{\odot}$. This means that most of the
dust mass resides in large grains (200:1). Using the canonical gas to
dust ratio of 100 the total disk mass becomes
2$\cdot$10$^{-4}$\,M$_{\odot}$. Note, that \ch{if the outer disk has a
lower value of $\Psi$,} a larger gas mass could be required to
reproduce the observations.

The results presented above naturally lead to the picture of a disk in
which the dust grains in the outer disk are coagulated and settled
towards the midplane, while the gas is still available to keep the PAH
molecules in the higher atmosphere of the disk. As already noted in
the theoretical study by \cite{2007A&A...473..457D}, growth and
settling of the dust grains leads to a natural increase of the
relative strength of the PAH signature, as is observed in this disk
and confirmed by our modeling effort.

\subsection{Context}
In general the study of \cite{2007A&A...473..457D} showed that the
natural outcome of a group I source after grain-growth and
sedimentation of the dust is a group II source, that maintains the
flaring structure for the gas. However, observational studies have
shown that most group II sources lack a flaring gas
distribution. \cite{2001A&A...365..476M}, \cite{2004A&A...422..621A}
\ch{and \cite{2010ApJ...PAHs}} showed that group I sources display
significantly more PAH emission. \cite{2005A&A...436..209A} showed
that group I sources have in general stronger [O\,I] emission.
Apparently the gas of most group II sources has either dramatically
decreased its scale height because of the lack of heating or the gas
has been dispersed from the disk. How disks can lose their gas is
currently being debated (see \citealt{2008PhST..130a4024H}), but
photoevaporation seems to be the most likely mechanism.

On the other hand there is a fair number of group II sources that do
show indications of a flaring gas distribution. Some group II sources
display PAH emission in their 10\,$\mu$m spectra, for example:
\object{HK\,Ori} (\citealt{2005A&A...437..189V}) and
\object{HD\,142666} (\ch{\citealt{2010A&A...HAEBE}}). Some group II
sources display the [O\,I]~6300\,\r{A} line in emission, for example:
\object{HD\,98922} (\citealt{2005A&A...436..209A}). \ch{And some show
both PAH and [O\,I] emission, for example: \object{HD\,101412}
(\citealt{2008A&A...491..809F}).}  These sources all seem to be in a
transitional phase from a gas rich flaring disk to a gas poor
self-shadowed disk. Roughly half of the known group II sources are in
this phase, which means we can infer that half of the life time of the
disk of a group II source is spent on the dispersal of the gas. An
\ch{estimate} for this time scale is the photoevaporation time scale,
which is on the order of $\sim$10$^6$\,yr
(\citealt{2009ApJ...690.1539G})

\section{Conclusions}
A comprehensive study was performed to map the distribution of the gas
and dust in the protoplanetary disk around HD\,95881. In \ch{the right
panel of Fig.\,\ref{fig:models}} we displayed a schematic
representation of the disk, which puts all results in perspective. The
AMBER K-band interferometry showed that there is an extended hot inner
region with emission coming from \ch{well} within the sublimation
radius. The detection of the [O\,I]~6300\,\r{A} indicated that the
disk has a flaring gas surface at large distances (from one to tens of
AU) from the star. The finding of PAH features in the Spitzer and
VISIR spectra confirmed the presence of an illuminated gas
surface. The resolved VISIR spectrum traced this surface up to 
\ch{radii of $\sim$100\,AU.}

We used the radiative transfer code MCMax
(\citealt{2009A&A...497..155M}) to create a model of the disk's
density and temperature structure. Our model satisfactorily reproduced
all of our observations. The main conclusions that followed from our
model are that the inner disk contains most of the \ch{visible} grains
and has a puffed up inner rim, the dust grains in the outer disk have
coagulated and settled towards the midplane, while the gas and PAH
mixture maintain a flaring geometry. Theory predicted the existence of
these type of disks (\citealt{2007A&A...473..457D}), while
observational trends showed that most of the sources with
self-shadowed dust distributions have dispersed \ch{the bulk of} their
gas. In this light HD\,95881 is a special source: it is in the
transition phase from a gas rich flaring dust disk to a gas poor
self-shadowed disk.

\begin{acknowledgements}
This research was sponsored by NWO under grant number
\ch{614.000.411}. M.  Min acknowledges financial support from the
Netherlands Organization for Scientific Research (NWO) through a Veni
grant. E. Pantin acknowledges financial support from the Agence
Nationale de la Recherche (ANR) of France through contract
ANR-07-BLAN-0221.
\end{acknowledgements}

\bibliographystyle{./aa.bst}
\bibliography{./12656bibl.bib}

\end{document}